\documentclass[reprint,prb,superscriptaddress]{revtex4-1}
 \usepackage{booktabs}
 \usepackage{amsmath,amssymb}
 \usepackage{graphicx}
 \usepackage{color}
 \usepackage{hyperref}
 \newcommand{\Heff}{H_\mathrm{eff}}
 \newcommand{\Hint}{H_\mathrm{int}}
 \newcommand{\Jeff}{J_\mathrm{eff}}
 \newcommand{\momentum}{\mathbf{k}}
 
 \newcommand{\U}{\mathcal{U}}
 \newcommand{\A}{\textstyle\frac{E\delta}{\omega}}
 \newcommand{\norm}[1]{\left\lVert#1\right\rVert}
 \newcommand{\bes}[1]{\mathcal{J}_{#1}}
 
\begin{document}

\title{Sublattice dynamics and quantum state transfer of doublons in 2D lattices}

\author{M. Bello}
\affiliation{Instituto de Ciencias de Materiales de Madrid, CSIC, E-28049, Spain}
\author{C. E. Creffield}
\affiliation{Departamento de F\'isica de Materiales, Universidad Complutense de Madrid, E-28040, Spain}
\author{G. Platero}
\affiliation{Instituto de Ciencias de Materiales de Madrid, CSIC, E-28049, Spain}
\date{\today}

\begin{abstract}
We analyze the dynamics of two strongly-interacting fermions moving in 2D lattices under the action 
of a periodic electric field, both with and without a magnetic flux. Due to the interaction, these particles 
bind together forming a doublon. We derive an effective Hamiltonian that allows us to understand the 
interplay between the interaction and the driving, revealing surprising effects that constrain the movement 
of the doublons. We show that it is possible to confine doublons to just the edges of the lattice, and also 
to a particular sublattice if different sites in the unit cell have different coordination numbers. Contrary to 
what happens in 1D systems, here we observe the coexistence of both topological and Shockley-like 
edge states when the system is in a non-trivial phase. 
\end{abstract}


\maketitle

\section{Introduction}
Tunneling dynamics of particles in lattices can be well understood with 
tight-binding Hamiltonians. In these models, quantum coherence is responsible 
for many exotic phenomena such as system revivals, quantum interference, and 
Rabi-like oscillations. It has now become possible to observe these effects in a 
variety of setups ranging from photonic crystals 
\cite{Lieb_photonic,Lieb_flatband,rhombi_photonic,photonicqw} to quantum dots 
\cite{petta,gaudreau,forster} and cold atoms trapped in optical lattices 
\cite{Bloch_Many-body_QGas,bloch,li,christiane}. In particular, quantum 
coherence allows the transfer of quantum information between different 
locations, a process known in the literature as quantum state transfer (QST). 
Given its importance in quantum information processing applications, QST has 
been the object of study in many experimental and theoretical works carried out 
in recent years \cite{rafa,bose,qst,sap}.
 
Adding a periodic driving potential considerably enriches the physics of these 
systems, and provides a means for controlling and manipulating them. Such 
driving can produce effects such as dynamical localization \cite{dunlap} and 
coherent destruction of tunneling (CDT) \cite{hanggi}, and can even be used to 
design artificial gauge fields \cite{goldman,cec_sols}. This flexibility and 
controllability makes driven lattice systems ideal for use as quantum simulators 
\cite{Struck_FrustratedMag,jotzu}. Despite the many advances in the field, 
however, driven \textit{interacting} systems have not been extensively studied 
yet. Understanding the role interactions play in these setups is a hard task of 
fundamental importance, however, since the behavior of the system may change 
drastically compared with the non-interacting case, and produce novel and 
unusual physics.

\begin{figure}[!ht]
 \includegraphics{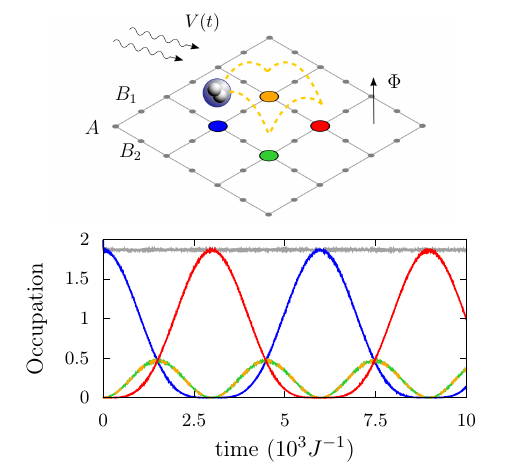}
 \caption{(Top) Scheme depicting a doublon propagating in a finite piece of the Lieb lattice under the effect 
 of an ac field and an external magnetic field perpendicular to the lattice. Using a periodic driving it is possible 
 to control doublon's effective hopping rate independently of the effective local chemical potential. Close to the 
 CDT condition, its dynamics becomes restricted to the particular sublattice where it was initially located 
 (colored sites), the occupation of the other sites being almost zero during the entire time evolution. 
 (Bottom) Example of sublattice dynamics. Time evolution is obtained by numerical integration using Hamiltonian 
 \eqref{Ht}, parameters: $\Phi=0$, $U=16J$, $\omega=2J$ and $E=4.8J/a$. The sum of the occupancies on the colored 
 sites (grey line) barely varies, staying close to 2.} \label{scheme}
\end{figure}

Our aim in this work is to extend QST to interacting systems of few-particles.
We investigate the dynamics of two strongly-interacting fermions in 2D 
lattices. The fermions can bind together repulsively, forming what is termed a 
``doublon'', a long-lived excitation whose decay is forbidden on energetic 
grounds \cite{winkler,Doublons_estability,Preiss2015}. Bosons can also bind 
together in this way, and the quantum walk for repulsively bound bosonic 
particles on a 1D lattice has been recently studied in \cite{compagno}. The 
regime of strongly interacting particles, i.e., that of doublons rather than
single particles, is interesting in itself. There are several experiments
analyzing the dynamics of high energy bound-states of ultracold fermions and
bosons \cite{Folling2007,schneider,Preiss2015}.

We derive an effective Hamiltonian describing the motion of single doublons in 
2D lattices coupled to circularly polarized ac fields, in the presence of a 
magnetic flux threading the lattice. For a special class of lattices, we 
demonstrate an interesting effect by which the doublon's dynamics is restricted 
to just one of the sublattices of the crystal. Not only that, our results show 
it is possible to confine the doublon dynamics to certain sites (those with 
same coordination number) on the edges of any finite system, and induce direct 
transfer of the doublon between distant sites, avoiding the intervening sites. 
This makes the process less susceptible to decoherence. Although we present 
results mainly for the Lieb lattice, the conclusions we draw apply to a wide 
set of 2D lattices.

\section{Model}
We consider a Hubbard model for fermions with an external ac field and a uniform 
magnetic field perpendicular to the plane of the lattice. The ac field couples 
to the particle density, and the magnetic flux induces phases in the hoppings 
such that the sum of the phases around a closed loop is the total flux threading 
the loop, measured in units of the magnetic flux quantum $\Phi_0 = h/e$. The 
system is then described by the tight-binding model:
\begin{multline}
	H(t) =  -J\sum\limits_{\langle i,j\rangle,\sigma} e^{i\phi_{ij}}c^\dagger_{i\sigma}c_{j\sigma}+
	U\sum\limits_i n_{i\uparrow}n_{i\downarrow} \\
          +\sum\limits_i V_i(t)(n_{i\uparrow}+n_{i\downarrow}) \label{Ht} \ ,
\end{multline}
where $c^\dagger_{i\sigma}$ ($c_{i\sigma}$) is for the creation (annihilation) 
operator of a fermion on site $i$ with spin $\sigma$, and 
$n_{i\sigma}=c^\dagger_{i\sigma}c_{i\sigma}$ is the usual number operator. We 
choose a circularly polarized driving: 
$V_i(t)=x_iE\cos(\omega t)+y_iE\sin(\omega t)$, where $x_i$ and $y_i$ are the 
coordinates of site $i$. The parameters of the model are the interaction 
strength $U$, the hopping amplitude $J$, and the ac field amplitude $E$ and 
frequency $\omega$.

In the strongly-interacting limit of the undriven model, particles can bind 
together repulsively forming a doublon \cite{charles_rhombi,charles_qdarray,
winkler}. This bound state consists of two particles with opposite spin occupying 
the same lattice site. If initially two particles form a doublon, they will 
remain bound together thereafter in the absence of dissipation. This can be 
understood on energetic grounds. The kinetic energy of a single particle in a 
lattice is limited by the width of the energy bands, which is proportional to the 
hopping amplitude; thus if $U\gg J$, doublons cannot decay into single particles, 
as energy would not be conserved. In this regime, the total double occupancy is 
approximately a conserved quantity, and one can obtain an effective Hamiltonian 
for doublons by means of a Schrieffer-Wolff transformation (SWT), projecting out 
single occupancy states \cite{hofmann}.

In the presence of an ac field one might expect the stability of doublons to be 
spoiled. To address this question we derive an effective Hamiltonian that 
includes both the interaction between particles and the periodic driving, using 
the so-called high frequency expansion (HFE). This method allows the effective 
Hamiltonian to be written as a power series in $1/\omega$, the different terms 
being functions of the Fourier components of the original time-periodic 
Hamiltonian \eqref{Ht} \cite{bukov}. A different effective Hamiltonian is 
obtained depending on whether the system is in the \textit{strongly-interacting 
regime} ($U\gg \omega>J$), or the \textit{high-frequency regime} ($\omega\gg 
U>J$). In the first case it corresponds to first performing a hopping 
renormalization and then the SWT, whereas in the second it is the other way 
around \cite{doublons_dimer,artificial_GF}. In the strongly-interacting regime, 
the driving can induce the formation and dissociation of doublons. These 
processes involve the absorption and emission of photons with a probability 
amplitude proportional to $\mathcal{J}_l(2E\delta/\omega)$ \cite{artificial_GF}. 
Thus, for small driving amplitudes, $2E\delta<\omega l = U$ (where $\delta$ is 
the distance between neighboring sites and $l$ the order of resonance), the 
probability is very small and doublons persist in time. Conversely, for high 
driving amplitudes the total double-occupancy of any given state changes 
considerably within a period. Up to first order, the effective Hamiltonian we 
find for the strongly-interacting regime with small driving amplitudes is  
\begin{equation}
	\Heff=\Jeff\sum\limits_{\langle i,j \rangle} e^{i2\phi_{ij}} d^\dagger_i d_j + \sum\limits_i \mu_i n^d_i \ ,
  \label{heff}
\end{equation}
where $d^\dagger_i=c^\dagger_{i\uparrow}c^\dagger_{i\downarrow}$ ($d_i$) is the 
creation (annihilation) operator of a doublon on site $i$, and 
$n^d_i=d^\dagger_id_i$ is the doublon number operator. In this result, additional 
terms including the interaction between doublons have been neglected, as we consider the dynamics of just a single doublon. Also, following the previous 
reasoning, we have neglected terms which correspond to transitions between 
single-occupancy and double-occupancy states caused by the driving. $\Jeff$ and 
$\mu_i$ can be written in terms of the original parameters as:
\begin{equation}
	\Jeff = \frac{2J^2}{U}\mathcal{J}_0\left(\frac{2E\delta}{\omega}\right) \ ,\quad
	\mu_i = \frac{2J^2}{U} z_i \ , \label{eff_param}
\end{equation}
where $z_i$ is the coordination number (the number of nearest neighbors) of site 
$i$. This dependence of the local effective chemical potential on the number of
neighbours comes from the second order process where the doublon splits, 
one of the particles remaining in the original site and the other one moving to one of its
neighbors, and then recombines again in the original site. The process may involve
any of the neighbors, so the total effect is an effective chemical potential
proportional to the coordination number. The effective hopping amplitude for the 
doublon is proportional to the zeroth-order Bessel function of the first kind, 
whose argument depends on the parameters of the ac field and the geometry of the 
lattice. This hopping renormalization is isotropic because the ac field 
polarization is circular. A generalization to other polarizations is 
straightforward but they lead to more complicated effective models. 

\section{Sublattice dynamics}
\begin{figure}
	\includegraphics{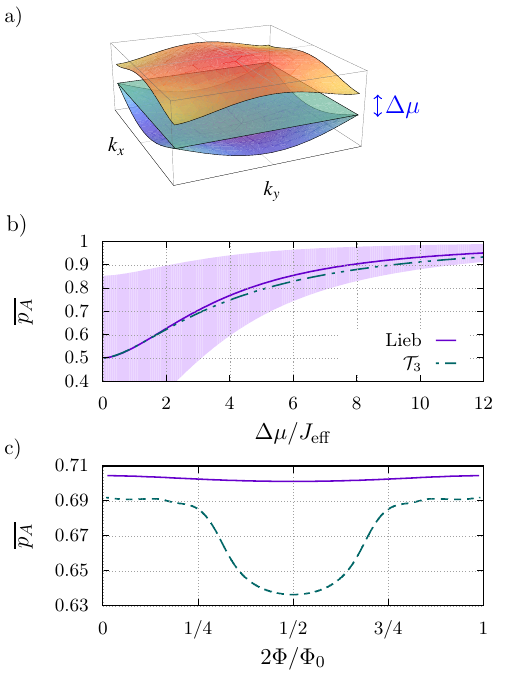}
	\caption{a) Energy bands for the Lieb lattice. The effective local potential 
	experienced by the doublon opens a gap in its energy spectrum. The ac driving allows
	the band width to be reduced, flattening the bands, while keeping the gap the same. This changes the relative 
	weight in sublattices $A$ and $B$ of the Bloch states corresponding to the upper and lower bands, 
	see Eq. \eqref{blochstates}.
	b) Calculation of the time averaged probability to remain in sublattice $A$ for the Lieb and the 
	$\mathcal{T}_3$ lattices with zero magnetic flux as a function of $r=\Delta\mu/\Jeff$. The light 
	purple area shows the value of $\sigma_A$ above and below the mean for the Lieb lattice. 
    Clearly as $\Delta\mu/\Jeff$ increases, the effectiveness of the sublattice confinement
    also grows.
    c) Graph of $\overline{p_A}$ as a function of the magnetic flux threading the elementary plaquette for $r=3$. 
	A much stronger dependence is observed for the $\mathcal{T}_3$ than for the
    Lieb lattice. When the magnetic flux is not zero, the calculation is more
    involved since it is necessary to take into account the larger magnetic unit cell.}
	\label{pa_energybands}
\end{figure}

As we can see in \eqref{eff_param}, the ac driving allows us to independently tune the effective hopping 
parameter with respect to the effective local potential. This has a big impact on the dynamics of doublons 
in lattices that can be divided into sublattices with different coordination numbers, such
as the Lieb lattice shown in  Fig. \ref{scheme}, and the $\mathcal{T}_3$ lattice \cite{chrisT3,vidal} 
shown in Fig. \ref{T3}. In both these examples, the effective Hamiltonian in
momentum space in the absence of an external magnetic flux can be expressed as 
$\Heff = \sum\limits_\momentum  \mathbf{\Psi^\dagger_k}\mathcal{H}(\momentum)\mathbf{\Psi_k}$, with 
\begin{equation}
	  \mathcal{H}(\momentum)=\begin{pmatrix}
                  \Delta\mu & f_1(\momentum) & f_2(\momentum) \\
                  f^*_1(\momentum) & 0 & 0\\
                  f^*_2(\momentum) & 0 & 0
                 \end{pmatrix} \ .
\end{equation}
$\mathbf{\Psi_k}=(d_{A,\momentum}, d_{B_1,\momentum}, d_{B_2,\momentum})^T$, 
$d_{A,\momentum}$ is the annihilation operator of a doublon with quasi-momentum
$\momentum$ in sublattice $A$; we define $d_{B_1,\momentum}$ and 
$d_{B_2,\momentum}$ analogously. Its eigenvalues and eigenvectors are:
\begin{align}
\epsilon_0 (\momentum) & =0 \ , \\
\epsilon_\pm(\momentum) & =\left(\Delta\mu \pm \sqrt{4|f_1(\momentum)|^2 + 4|f_2(\momentum)|^2 + \Delta\mu^2}\right)/2 \ . \\
\lvert{u^0_\momentum}\rangle & = \frac{1}{N}\left( 0, -\frac{f_1(\momentum)}{f_2(\momentum)}, 1\right) \ , \\
\lvert{u^\pm_\momentum}\rangle & = \frac{1}{N}\left( \frac{\epsilon_\pm(\momentum)}{f^*_2(\momentum)}, \frac{f^*_1(\momentum)}{f^*_2(\momentum)}, 1\right) \ .
\label{blochstates}
\end{align}
Here $N$ is just a normalization constant. Note how the states of the flat band do not have weight on 
the $A$ sites of the lattice \cite{Lieb_flatband}. We present the energy bands for the Lieb lattice
in Fig. \ref{pa_energybands}a, which clearly shows the band splitting produced by 
the chemical potential difference between the two sublattices, $\Delta\mu=2J^2(z_A-z_B)/U$. 
The functions $f_1$ and $f_2$ depend on the particular lattice geometry, as shown in Table \ref{table}. 
They are proportional to $\Jeff$, which can be tuned by the ac driving. 
In particular, the relative weight on the $A$ sublattice of the Bloch states corresponding to the upper 
(lower) band can be increased (reduced) by tuning the ac field parameters closer to the CDT condition.

\begin{table*}
	\begin{ruledtabular}
	\begin{tabular}{ l c c }
		  & $f_1(\mathbf{k})$ & $f_2(\mathbf{k})$\\ \hline
		$\mathcal{T}_3$ 
		& $\Jeff\left[e^{-i(k_x+k_y/\sqrt{3})a/2}+e^{i(k_x-k_y\sqrt{3})a/2}+e^{ik_ya/\sqrt{3}}\right]$ 
		& $f_2(\mathbf{k})=f^*_1(\mathbf{k})$\\
		Lieb & $2\Jeff\cos(k_xa/2)$ & $2\Jeff\cos(k_ya/2)$
	\end{tabular}
	\end{ruledtabular}
	\caption{Functions characterizing the energy bands of the Lieb and $\mathcal{T}_3$ lattices.}
	\label{table}
\end{table*}

When studying \textit{quantum walks} \cite{ctqw}, i.e.\ the coherent evolution of particles in networks, it is 
natural to ask about the probability of finding a particle that was initially on site $i$, to be on site $j$ after 
a certain time $t$, that is $p_{ij}(t)=|\langle i |U(t)| j \rangle |^2=|\langle i |e^{-iH t}| j \rangle |^2$. 
Using \eqref{heff} as the effective single-particle Hamiltonian for the doublon, we define 
$p_A(t)=\frac{1}{N_A}\sum_{i,j\in A} p_{ij}(t)$, which is the probability for the doublon to 
remain in sublattice $A$ at time $t$. To demonstrate sublattice confinement, we can compute the long time 
average $\overline{p_A}$ and variance $\sigma^2_A=(\overline{{p_A}^2}-\overline{p_A}^2)$, 
Fig. \ref{pa_energybands}b, see appendix B. Their values are mainly determined by the ratio: $r=\Delta\mu/\Jeff$. 
As shown in Fig. \ref{pa_energybands}b, the probability $\overline{p_A}$ can be enhanced by tuning $r$ to 
larger values, meaning that it is possible to confine the doublon's dynamics to a single sublattice by suitably 
changing the ac field parameters (see Eq.\ \eqref{eff_param}). The variance of this average 
probability also reduces when going in this direction. We have also computed the dependence of $\overline{p_A}$ 
with the magnetic flux threading the unit cell, see Fig. \ref{pa_energybands}c; however, its variation 
turns out to be minor, with $\overline{p_A}$ gently increasing as the flux is tuned away from $2 \Phi /\Phi_0 = 1/2$. 
A much stronger dependence is observed for the $\mathcal{T}_3$ than for the Lieb lattice. This is to be expected, 
as Aharonov-Bohm phases have more dramatic effects in the  $\mathcal{T}_3$ lattice, notably the caging effect 
that occurs for a magnetic flux $\Phi / \Phi_0 = 1/2$ in the single-particle case \cite{vidal,abc_interaction,charles_rhombi}.

\begin{figure}[!htb]
    \centering\includegraphics{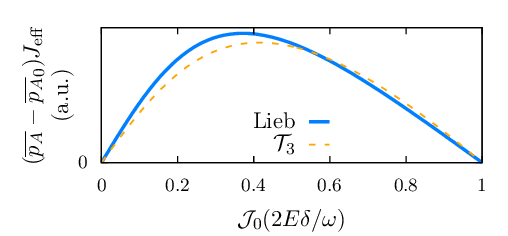}
    \caption{Sublattice dynamics optimization for the Lieb and the
    $\mathcal{T}_3$ lattices. $\overline{p_A}_0\equiv\overline{p_A}\rvert_{E=0}$, 
is the average probability to remain in sublattice A when there is no ac
driving.}
    \label{p_a_optimize}
\end{figure}

\begin{figure}[!htbp]
    \includegraphics{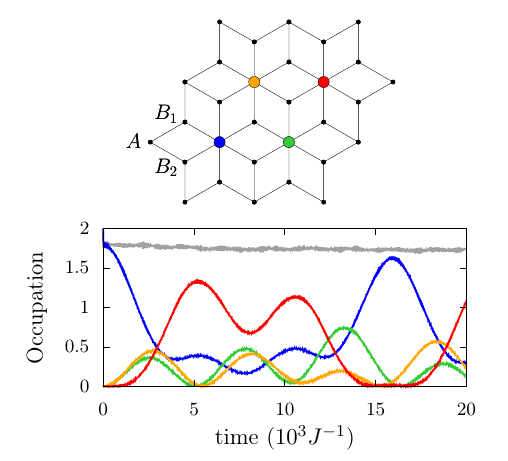}
	\caption{Time evolution for a finite piece of the $\mathcal{T}_3$ lattice (see upper scheme). The parameters 
	of the system are: $U=16J$, $\Phi=0$, $\omega=2J$ and $E=4.2J/a$. The initial condition is two electrons with 
	opposite spin occupying the lower-left site (blue). The average occupation on the sites not shown never exceeds 
	the value of 0.02 per site. The doublon mostly remains in the sublattice where it was initially located. 
	The grey line shows the sum of the occupancies of the four colored sites of the scheme.}
	\label{T3}
\end{figure}

From this analysis, it is clear that by tuning the ac field parameters closer to the CDT condition (i.e.\ when $2E\delta/\omega$ is a zero of $\mathcal{J}_0$) 
one can enhance the confinement to the sublattice, at the expense of slowing down 
the dynamics. In Fig. \ref{p_a_optimize} we plot the quantity
$\left(\overline{p_A}-\overline{p_A}\rvert_{E=0}\right)\Jeff$
which corresponds to the difference of the average probability to remain in 
sublattice A in the driven and undriven case multiplied by the effective doublon 
hopping as a function of the hopping renormalization. It gives an idea of the 
optimal parameters regime for having sublattice localization while keeping the 
time-scales in which dynamics take place finite enough to observe it in 
experiment. Examples of sublattice dynamics are shown in figures \ref{scheme} and 
\ref{T3}. 

\section{Edge dynamics and QST}
\begin{figure}[ht]
    \includegraphics{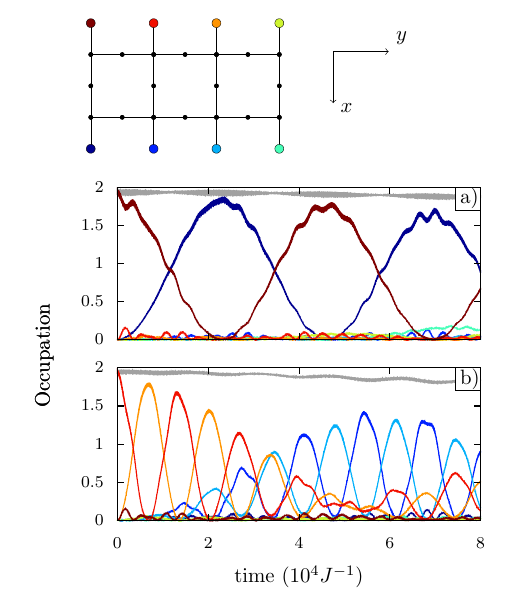}
	\caption{Time evolution for a finite piece of the Lieb lattice, shown in the upper panel.
	$U=16J$, $\omega=2J$ and $E=4J/a$. 
    a) Initially the doublon occupies the top-left site, in brown. The doublon oscillates between the top-left and 
	bottom-left sites of the lattice. This oscillation is transferred to the right-top and bottom sites over a longer time 
	period. b) The initial condition is now a doublon occupying the middle-top site (in red). The doublon propagates 
	mainly through the top and bottom edges. It performs oscillations between the middle sites 
	and on a larger timescale it is transferred to the opposite edge without occupying the intermediate sites in the bulk; 
	this is a long-range transfer process. The occupation on the intervening sites never exceeds a value of 0.012 per site. 
	The grey line gives the sum of the occupancies in all eight sites of the top and bottom edges.}
	\label{Lieb_edge}
\end{figure}

\begin{figure}[!htb]
    \centering\includegraphics{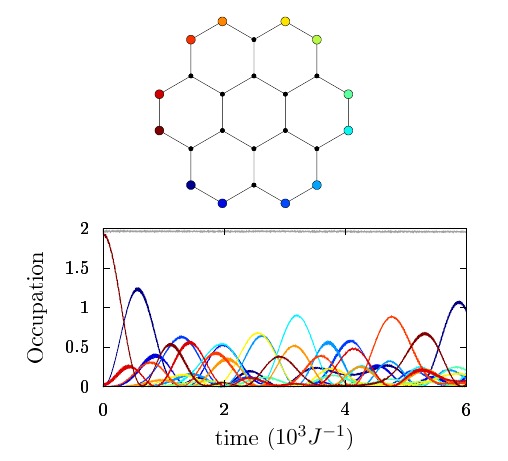}
    \caption{Time evolution for a finite piece of the honeycomb lattice (see upper scheme). The parameters 
	of the system are: $U=16J$, $\Phi=0$, $\omega=2J$ and $E=4.2J/a$. The doublon is initially occupying the 
    edge site in brown. It propagates mainly through the sites at the edge with only two neighbours. 
    The average occupation on the sites not shown never exceeds the value of 0.005 per site. In grey, 
	the total occupancy on the edge (marked sites).}
    \label{graphene_edge}
\end{figure}

In light of the effective Hamiltonian we have derived, new effects particular to systems with boundaries can be predicted. 
The sites on the edges of a finite lattice necessarily have fewer neighbors than those in the bulk and therefore have 
a smaller chemical potential (Eq. \eqref{eff_param}). This produces eigenstates localized on the edges, 
which are of the usual Shockley or Tamm type. As a consequence the doublon's dynamics can be confined to just the edges. 
We show an example of edge confinement in Fig. \ref{Lieb_edge}. Importantly, this effect is general in the sense that 
it happens in {\em any} kind of lattice, see Fig. \ref{graphene_edge}. The resulting dynamics strongly depends on the particular shape of 
the boundary and the initial condition, as different sites of the edge can have different number of neighbors. In some 
cases the direct transfer of doublons between distant sites of the boundary can happen, as shown in Fig. \ref{Lieb_edge}. 
This occurs via the hybridization of the edge states on opposite edges, forming bi-localized eigenstates that give rise 
to Rabi-like oscillations. The transfer time increases exponentially with the number of sites that separate one edge 
from the other.

\section{Topological edge states for doublons}
When comparing our effective model \eqref{heff} to that corresponding to a Chern insulator, the only difference is the local 
chemical potential term \cite{Bernevig, tenfold_way}. It is well known that strong disorder potentials eventually destroy the 
topological properties of Chern insulators, as they transition to a trivial Anderson insulator by a mechanism known as 
``levitation and annihilation'' of extended states \cite{vozmediano,Anderson_localization}. Nonetheless, the chemical potential 
term \eqref{eff_param} constitutes a very particular form of disorder that does not affect the topology of the system. This is 
in contrast to the much more drastic effect it has in 1D topological models, such as the SSH model, where it breaks the 
particle-hole symmetry needed to obtain a phase other than the trivial one \cite{doublons_dimer}. 

\begin{figure}
    \centering
    \includegraphics{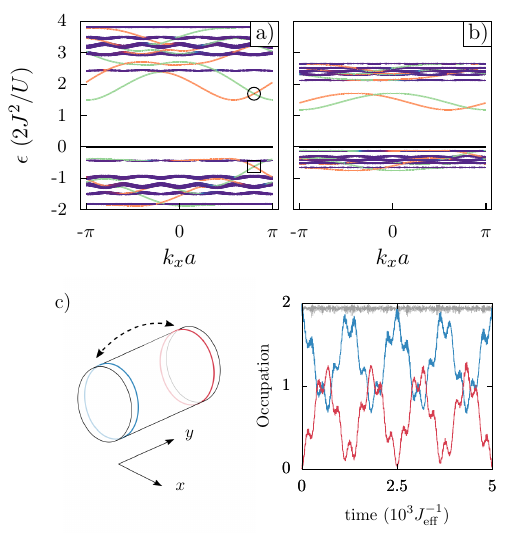}
	\caption{a) Energy spectrum of the undriven effective Hamiltonian for a doublon in a Lieb lattice ribbon 
	with $N_y=100$ unit cells in the $y$ direction. $\Phi/\Phi_0=1/10$. The 
	color indicates the average position in the $y$ direction of the eigenstate. Orange and green 
	correspond to localized states in the left and right edge respectively, whereas purple corresponds to extended 
	bulk states. At $k_x=k_0=4\pi/5$ two anticrossings between edge states are marked 
	($\square$ and $\bigcirc$). The hybridization gap decays exponentially with the number of 
	cells, $N_y$, and it is barely noticeable for the case shown here. b) Energy spectrum for 
	$\mathcal{J}_0(2E\delta/\omega)=0.5$. As the bands shrink 
	due to the effective hopping renormalization, two bands of Shockley-like edge states (one for each 
	edge) separate from the upper part of the spectrum.
	c) Time evolution of the occupation in the left and right edges of a thin ribbon, or cylinder, of the 
	Lieb lattice, with $N_y=5$ unit cells in the $y$ direction. The magnetic flux is $\Phi=\Phi_0/10$ and there is no 
	ac driving. The momentum of the initial state is set to $k_0$, with a probability homogeneously distributed among 
	the sites at the left edge (blue in the scheme), and zero elsewhere. In red, the occupancy of the sites at the 
	right edge. The sum of the occupancies on both edges (in grey) remains constant and close to 2, indicating that 
	the doublon does not occupy sites in the bulk of the lattice.}
	\label{cylinder_spectrum}
\end{figure}

To analyze the effect of the magnetic flux on the doublons dynamics, we choose a vector potential $\mathbf{A}=B y \mathbf{u}_x$ 
corresponding to the Landau gauge, and study a system periodic in the $x$ direction but finite in $y$. 
Interestingly, we observe in  Fig. \ref{cylinder_spectrum}a,b that the energy spectrum shows the coexistence 
of both chiral topological edge states and non-chiral Shockley-like edge states.
 
Furthermore, in narrow ribbons the topological edge states can also hybridize, enabling the transfer of the 
doublon between the two edges of the ribbon, as we saw previously for the Shockley edge states. Looking 
at the energy spectrum, we can observe values of $k_x=k_0$ for which there are anticrossings between 
the edge states. At those values of momentum, a probability density initially peaked around 
one of the edges of the ribbon will oscillate between the two edges, while being almost equal to zero in the 
bulk, see Fig. \ref{cylinder_spectrum}c .

In this work we have concentrated mainly on the effect of the circular shaking
term, $V(t)$, with the magnetic flux taken as a given. A variety of techniques
now exist to produce the Peierls phases in cold-atom experiments, such as the
photon-assisted tunneling schemes described in Refs.
\onlinecite{jotzu,schneider}, the implementations based on assisted Raman transitions used by the Bloch and Ketterle groups 
\onlinecite{Aidelsburger2013,Miyake2013}, or by using Berry phases 
to mimic the Peierls phases. One exciting possibility would be to produce the
Peierls phases also by shaking, so that the entire effective Hamiltonian would
be produced by periodic driving. Early works on generating gauge fields on a
lattice via periodic shaking were
restricted to producing staggered fluxes on triangular lattices 
\cite{Struck_FrustratedMag}. Obtaining a \textit{uniform} field (of the type that 
we require in our system) on a lattice in which the plaquettes have parallel 
sides, such as the square lattice and Lieb lattice, is a much more involved 
problem, which requires special treatment such as ``split-driving'' \cite{cec_sols}.

\section{Conclusions}
We have analyzed the dynamics of two strongly interacting fermions in 2D 
lattices. A special property of the doublon is that it experiences a local 
chemical potential that depends on the coordination number of the lattice site.  
We propose the use of an ac driving to independently tune the doublon effective 
hopping and this local chemical potential. If a lattice contains a sublattice of 
sites with a certain coordination number, different from the coordination number 
of the remaining sites, this effect can be harnessed to limit the propagation of 
the doublon to just that sublattice. In finite samples this effect can also be 
used to confine the doublon to particular sites at the edges. We also discuss 
the coexistence of topological and Shockley edge states in 2D systems threaded 
by a magnetic flux. This coexistence, which does not occur in 1D systems with non-trivial topology, allows the direct doublon transfer between edges in a 
richer manner than in 1D systems, via the coherent superpositions of either 
Shockley or topological edge states. Our analysis is valid for any 2D lattice 
and can be experimentally investigated in cold atom lattices 
\cite{Taie_experimentLieb} or photonic crystals \cite{doublons_1DPC}. Developing
this work to address a many-particle scenario is an exciting future avenue for
research. However, even the two-particle results we report could be of relevance
to experimentalists, as these effects could be used to distinguish single
particles, as opposed to doublons, in a dilute gas just by looking at its
dynamics.

\acknowledgments
We would like to thank Alvaro G\'omez-Le\'on for enlightening discussions. 
MB and GP were supported by Spain's MINECO through Grant No. MAT2014-58241-P, and
CEC by Grant No. FIS2013-41716-P. 

\appendix
\section{Effective Hamiltonian for doublons}
We start from a Fermi-Hubbard model with an ac field that couples to the particle density and 
a magnetic flux that induces complex phases in the hoppings. The Hamiltonian of the system is
\begin{multline}
	H(t)=-J\sum\limits_{\langle i,j \rangle, \ \sigma} e^{i\phi_{ji}} c^\dagger_{j\sigma}c_{i\sigma}
	+ U\sum\limits_i n_{i\uparrow}n_{i\downarrow} + \\ \sum\limits_{i}V_i(t)(n_{i\uparrow}+n_{i\downarrow})
	\equiv H_J + H_U + H_{AC}(t) \ . \label{original}
\end{multline}
For a time-periodic Hamiltonian, $H(t+T)=H(t)$ with $T=2\pi/\omega$, Floquet's theorem 
permits us to write the time-evolution operator $U(t_2,t_1)$ as
\begin{equation}
	U(t_2,t_1)=e^{-iK(t_2)}e^{-i\Heff(t_2-t_1)}e^{iK(t_1)} \ ,
\end{equation}
where $\Heff$ is a time independent (effective) Hamiltonian and $K(t)$ is a $T$-periodic 
self-adjoint operator. $\Heff$ governs the long-term dynamics whereas $e^{-iK(t)}$, 
also known as the \textit{micromotion operator}, accounts for the fast dynamics 
occurring within a period. Following several perturbative methods 
\cite{andre_HFE,mikami_HFE}, it is possible to find expressions for these operators 
as power series in $1/\omega$
\begin{equation}
	\Heff=\sum\limits_{n=0}^\infty \frac{H^{[n]}}{\omega^n} \ , \quad 
	K(t)=\sum\limits_{n=0}^\infty \frac{K^{[n]}(t)}{\omega^n} \ .
\end{equation}
The different terms in these expansions have a progressively more complicated 
dependence on the Fourier components of the original Hamiltonian, 
${H^{(q)}=T^{-1}\int_0^T H(t)e^{i\omega q t}dt}$. The 
first three of them are:
\begin{align}
	H^{[0]} & =H^{(0)} \ , \quad
	H^{[1]} = \sum\limits_{q\neq 0}\frac{H^{(-q)}H^{(q)}}{q} \ , \\
	H^{[2]} & = \sum\limits_{q, p\neq 0} \left( \frac{H^{(-q)}H^{(q-p)}H^{(p )}}{q p} -\frac{H^{(-q)}H^{(q)}H^{(0)}}{q^2} \right)
\end{align}

\begin{figure}
	\centering
	\includegraphics{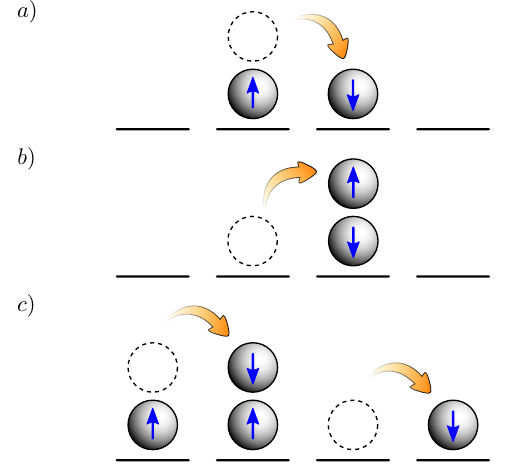}
	\caption{Schematic representation of the different hoppings: 
	a) $h^-_{ij\sigma}$, b) $h^+_{ij\sigma}$ and c) $h^0_{ij\sigma}$.}
	\label{hoppings}
\end{figure}

Before deriving the effective Hamiltonian, it is convenient to transform the original Hamiltonian 
\eqref{original} into the rotating frame with respect to both the interaction and the ac field
\begin{align}
	& \Hint(t)=\U^\dagger(t)H(t)\U(t)-i\U^\dagger(t)\partial_t\U(t) \ , \\
	& \U(t)=e^{-iH_Ut-i\int H_{AC}(t)dt} \ .
\end{align}
It can be written as:
\begin{align}
	\Hint(t) & = -\sum\limits_{\langle i,j \rangle, \ \sigma} Je^{i\mathbf{A}(t)\cdot \mathbf{d}_{ij}}
	\left[1-n_{i\overline{\sigma}}\left(1-e^{iUt}\right)\right]\times \nonumber \\ 
	& \qquad\qquad e^{i\phi_{ij}} c^\dagger_{i\sigma}c_{j\sigma} \left[1-n_{j\overline{\sigma}}\left(1-e^{-iUt}\right)\right] \\
	& = -\sum\limits_{\langle i,j \rangle, \ \sigma} Je^{i(\mathbf{A}(t)\cdot \mathbf{d}_{ij} + \phi_{ij})} 
	 \left[ h^0_{ij\sigma} \right. \nonumber \\ 
	& \qquad\qquad\qquad \left. + e^{iUt}h^+_{ij\sigma} + e^{-iUt}h^-_{ij\sigma}\right] \ .
\end{align}
Here, we have defined: 
\begin{align}
	h^0_{ij\sigma} & =n_{i\overline{\sigma}}c^\dagger_{i\sigma}c_{j\sigma}n_{j\overline{\sigma}} 
	+(1-n_{i\overline{\sigma}})c^\dagger_{i\sigma}c_{j\sigma}(1-n_{j\overline{\sigma}}) \ , \label{h0} \\
	h^+_{ij\sigma} & =n_{i\overline{\sigma}}c^\dagger_{i\sigma}c_{j\sigma}(1-n_{j\overline{\sigma}}) \ , \label{h+} \\
	h^-_{ij\sigma} & =(h^+_{ji\sigma})^\dagger= 
	(1-n_{i\overline{\sigma}})c^\dagger_{i\sigma}c_{j\sigma}n_{j\overline{\sigma}} \ . \label{h-}
\end{align}
The operators $h^0_{ij\sigma}$ involve hopping processes that conserve the total double occupancy, 
while $h^+_{ij\sigma}$ and $h^-_{ij\sigma}$ raise and lower the total double occupancy respectively 
(see Fig. \ref{hoppings}). 
$\mathbf{A}(t)$ is a vector potential that corresponds to the ac field. In the case of circular 
polarization: $\mathbf{A}(t)=(\cos\omega t, \sin\omega t)E/\omega$; 
$\mathbf{d}_{ij}=\mathbf{r}_i-\mathbf{r}_j$ is the vector connecting sites $i$ and $j$. In order 
to apply the HFE we need to find a common frequency. We will consider first the resonant regime, 
$U=l\omega$, and then, by means of analytical continuation, obtain the \textit{strongly-interacting limit} 
($U\gg\omega>J$) and the \textit{high-frequency limit} ($\omega\gg U >J$). The Fourier components of 
$\Hint(t)$ are
\begin{equation}
	\Hint^{(q)}=-\sum\limits_{\langle i,j \rangle, \ \sigma} J^{(q)}_{ij}h^0_{ij\sigma}+
	J^{(q+l)}_{ij}h^+_{ij\sigma}+J^{(q-l)}_{ij}h^-_{ij\sigma} \ ,
\end{equation}
where (using the Jacobi-Anger identity)
\begin{align}
	J^{(q)}_{ij} & = J e^{i\phi_{ij}} e^{-iq\alpha_{ij}}\bes{q}(\A) \nonumber \\
	J^{(q)}_{ji} & = \left[J^{(-q)}_{ij}\right]^*= J e^{-i\phi_{ij}} e^{-iq\alpha_{ij}}\bes{-q}(\A) \ , 
\end{align}
with $\alpha_{ij}=\arctan(d^y_{ij}/d^x_{ij})$ and $\delta=|\mathbf{d}_{ij}|$. $\bes{q}$ stands for 
the Bessel function of first kind of order $q$.

Now, the zeroth-order approximation in the HFE is given by:
\begin{multline}
	\Hint^{[0]} =- J\sum\limits_{\langle i,j \rangle, \ \sigma} e^{i\phi_{ij}} \left[
	\bes{0}\left(\A\right) h^0_{ij\sigma}+ \right.\\
	\left. \bes{l}\left(\A\right)h^+_{ij\sigma}+\bes{-l}\left(\A\right)h^-_{ij\sigma} \right] \ .
\end{multline}
In contrast to the undriven case, the total double occupancy is not an approximate conserved quantity 
in the strongly interacting limit. There are terms proportional to $\bes{l}(\A)$ that correspond to 
the formation and dissociation of doublons assisted by the ac field. However, for low driving amplitudes 
($E\delta/\omega < l$) the probability for these processes to occur is very small and we can neglect them. 
It is in this low amplitude regime where it makes sense to consider an effective Hamiltonian for the 
double-occupancy sector of the space of states. Thus, we will ignore the terms that go with $h^0_{ij\sigma}$ 
because they act non-trivially only on states with some single-occupancy. 

In the next order of the HFE, there will appear more terms that do not conserve the total double 
occupancy, which we neglect, and from those which do conserve it, we only keep the ones that act 
on the doublon's subspace of states:
\begin{multline}
	\frac{\Hint^{[1]}}{\omega} 
	\simeq \sum\limits_{\langle i, j \rangle, \ \sigma}\left[ \frac{1}{\omega}\sum\limits_{q\neq 0} 
         \frac{\bes{-q+l}^2\left(\A\right)J^2 h^+_{ij\sigma}h^-_{ji\sigma}}{q} \right. \\ 
          + \left. \frac{1}{\omega}\sum\limits_{q\neq 0}
         \frac{\bes{-q+l}\left(\A\right)\bes{q-l}\left(\A\right)J^2 
         e^{i2\phi_{ij}} h^+_{ij\sigma}h^-_{ij\overline\sigma}}{q}\right] \ .
\end{multline}
Here, the first term is equal to:
\begin{multline}
  \frac{J^2}{\omega}\sum\limits_{p\neq -l} \frac{\bes{p}^2\left(\A\right)h^+_{ij\sigma}h^-_{ji\sigma}}{l-p}= \\
	\frac{J^2}{U}\sum\limits_{p\neq -l} \frac{\bes{p}^2\left(\A\right)}{1-p\omega/U}
	\left(n_{i\overline\sigma}n_{i\sigma}-n_{i\overline\sigma}n_{i\sigma}n_{j\sigma}n_{j\overline\sigma}\right) \ ,
\end{multline}
and the second term is equal to:
\begin{multline}
	\frac{J^2e^{i2\phi_{ij}}}{\omega}\sum\limits_{p\neq -l} 
	\frac{\bes{p}\left(\A\right)\bes{-p}\left(\A\right)h^+_{ij\sigma}h^-_{ij\overline\sigma}}{l-p}=\\
	\frac{J^2e^{i2\phi_{ij}}}{U}\sum\limits_{p\neq -l} \frac{\bes{p}\left(\A\right)\bes{-p}\left(\A\right)}{1-p\omega/U}
	c^\dagger_{i\sigma} c^{\dagger}_{i\overline\sigma} c_{j\overline\sigma} c_{j \sigma} \ .
\end{multline}
In the limit $U\gg\omega >J$, $p\omega/U\ll 1$ and we can approximate all the denominators in the above expressions as 1. 
Also, when analytically continuing the formulas for values of $U$ other than multiples of the frequency, the restriction 
$p\neq -l$ has no meaning. Finally, using the identities 
\begin{align}
	&\sum\limits_{q=-\infty}^\infty \bes{q}^2(\alpha)=1 \ , \\
	&\sum\limits_{q=-\infty}^\infty \bes{q}(\alpha)\bes{k-q}(\beta)=\bes{k}(\alpha+\beta) \ ;
\end{align}
we arrive at
\begin{align}
	& \Heff^{U\gg\omega}=\Jeff\sum\limits_{\langle i,j \rangle} 
	e^{i2\phi_{ij}} d^\dagger_i d_j + \sum\limits_i \mu_i n^d_i
	-\frac{2J^2}{U}\sum\limits_{\langle i,j \rangle} n^d_i n^d_j  \ , \label{effective}\\
	& \Jeff\equiv2J^2\bes{0}\left(2\A\right)/U\ , \ \mu_i\equiv 2J^2z_i/U \ . 
\end{align}
Here we have expressed the effective Hamiltonian in terms of the doublon creation and annihilation operators, 
$d^\dagger_i=c^\dagger_{i\uparrow}c^\dagger_{i\downarrow}$ and $d_i=c_{i\downarrow}c_{i\uparrow}$, and the 
doublon number operator $n^d_i=d^\dagger_i d_i$; $z_i$ is the number of neighbours of site $i$. Importantly, 
there is a term that corresponds to the attractive interaction between neighboring doublons, but since we 
only have one doublon in the system, we do not take it into account.

For completeness we give also the result in the other limit: $\omega\gg U >J$. Now $p\omega/U$ is very large and 
all the terms in the sums are very small except those for $p=0$. The effective Hamiltonian in this case would be:
\begin{align}
	& \Heff^{\omega\gg U}=\Jeff\sum\limits_{\langle i,j \rangle} e^{i2\phi_{ij}} 
	d^\dagger_i d_j + \sum\limits_i \mu_i n^d_i
	-\Jeff\sum\limits_{\langle i,j \rangle} n^d_i n^d_j  \ ,\\
	& \Jeff\equiv2J^2\bes{0}^2\left(\A\right)/U\ , \ \mu_i\equiv \Jeff z_i \ .
\end{align}
It is worth mentioning that these results could also be obtained by applying the HFE sequentially, integrating first the 
fast varying terms corresponding to the leading energy scale in the system \cite{artificial_GF}. We also note that 
higher order corrections will include complex next-nearest-neighbor hoppings that break the time-reversal symmetry 
in systems without the presence of a magnetic flux. Nonetheless, we expect them not to be very significant for the 
effects of sublattice and edge confinement discussed in the main text. 
 
\section{Time average and standard deviation}
According to the definition, the probability $p_A(t)$ is $p_A(t)=\norm{U_A(t)}^2/N_A$,
where $\norm{\cdot}$ denotes the HilbertÐSchmidt norm, and $U_A(t)=P_AU(t)P_A$ is the time-evolution 
operator projected on the subspace of the $A$ sublattice. Using the spectral decomposition, 
\begin{equation}
	U(t) =\sum\limits_{\momentum}\sum\limits_n e^{-i\epsilon_n(\momentum)t}\lvert{u^n_\momentum}\rangle\langle{u^n_\momentum}\lvert \ ,
	\quad n\in\{0,\pm\} \ ;
\end{equation}
we can express
\begin{align}
	 \norm{U_A (t)}^2 & =\sum\limits_\momentum \left\lvert\frac{e^{-i\epsilon_+(\momentum)t}}{1+g_+(\momentum)}+\frac{e^{-i\epsilon_-(\momentum)t}}{1+g_-(\momentum)}\right\rvert^2\\
  & = \sum\limits_\momentum \left[\frac{1}{1+g_+(\momentum)}\right]^2+\left[\frac{1}{1+g_-(\momentum)}\right]^2 \nonumber\\
  &\qquad +\frac{2\cos(\epsilon_+t-\epsilon_-t)}{\left(1+g_+(\momentum)\right)\left(1+g_-(\momentum)\right)} \ ,
\end{align}
where we have defined $g_+(\momentum)=\frac{|f_1(\momentum)|^2+|f_2(\momentum)|^2}{\epsilon^2_+(\momentum)}$ and 
$g_-(\momentum)=\frac{|f_1(\momentum)|^2+|f_2(\momentum)|^2}{\epsilon^2_-(\momentum)}$. The time average is given by
\begin{align}
	\overline{p_A} & =\lim_{t\rightarrow\infty}\frac{1}{t}\int_0^t p_A(t')dt' \\
	& \simeq \frac{1}{V}\int_\mathrm{FBZ} \left[\frac{1}{1+g_+(\momentum)}\right]^2 + \left[\frac{1}{1+g_-(\momentum)}\right]^2 d\momentum \ .
\end{align}
Here $V$ stands for the area of the first Brillouin zone (FBZ). The value of this integral as a function of 
$r=\Delta\mu/\Jeff$ is shown in Fig. 2 in the main article. In a similar way we can compute the variance of 
$p_A$ as
\begin{align}
	\sigma^2_A & =\overline{{p_A}^2}-\overline{p_A}^2 \ , \quad\text{with} \\
	\overline{p_A}^2 & = \lim_{t\rightarrow\infty}\frac{1}{t}\int_0^t [p_A(t')]^2 dt' \ ,\\
	\sigma^2_A & \simeq \frac{1}{V^2}\int_\mathrm{FBZ} \frac{2}{(1+g_+(\momentum))^2(1+g_-(\momentum))^2} d\momentum \ .
\end{align}

When the magnetic flux is not zero the calculation is more involved since it is necessary to take into 
account the larger magnetic unit cell. 

\bibliography{doublons2D}
\end{document}